\documentclass[12pt]{article}

\begin{document}
\def\bib#1{[{\ref{#1}}]}

\title{  Cosmological Black Holes}
\author{Cosimo Stornaiolo
$^{1,2}$\thanks{e-mail:cosmo@na.infn.it}\\
{\em $^{1}$\small Istituto Nazionale di Fisica
Nucleare,} {\em \small Sezione di Napoli,}\\ {\em \small Complesso
Universitario di Monte S. Angelo}\\ {\em \small Edificio N', via
Cinthia, 45 -- 80126 Napoli}\\ {\em $~^{2}$\small Dipartimento di
Scienze Fisiche,} {\em \small Universit\`{a} degli Studi
``Federico II'' di Napoli,}\\ {\em \small Complesso Universitario
di Monte S. Angelo}\\ {\em\small Edificio N', via Cinthia, 45 --
80126 Napoli}
  }

\maketitle

\begin{abstract}
In this paper we propose a model for the formation of the
cosmological voids. We show that cosmological voids can form
directly after the collapse of extremely large wavelength
perturbations into low-density black holes or cosmological black
holes (CBH). Consequently the voids are formed by the comoving
expansion of the matter that surrounds the collapsed perturbation.
It follows that the universe evolves, in first approximation,
according to the Einstein-Straus cosmological model. We discuss
finally the possibility to detect the presence of these black
holes through their weak and strong lensing effects and their
influence on the cosmic background radiation.
\end{abstract}

 key words cosmology: theory, dark matter, black hole physics

\vskip 0.5truecm

\section{\normalsize Introduction}
One of the most intriguing features of the universe is that
galaxies tend to lie on sheet-like structures surrounding voids
with typical sizes of about $40-50h^{-1}$ Mpc \bib{peacock}.

The existence of voids has been  evident after the discovery by
Kirshner et al. of a $60$ Mpc large void in the  B\"{o}otes
constellation \bib{kirsh}. Systematic surveys have shown the
existence of many regions with similar characteristics
\bib{delapparent}, \bib{geller}. Today it is believed that voids
occupy about 50 per cent of the volume of the universe (e.g. see
\bib{El-Ad:1997af}).

From the observational point of view, one of the most important
issues is whether the voids are or are not really empty regions.
IRAS surveys indicate  the absence of infra-red galaxies, other
considerations lead  to exclude the presence of dark matter inside
them (see references in \bib{El-Ad:1997af}).

But as observed by Peebles in \bib{peebvoid}  the low dispersion
of the velocities of the galaxies indicates, when $\Omega_{m}=1$,
that most of the matter must be inside the voids, but the author
suggests that this could also be true even in the case of small
$\Omega_{m}$. Moreover, very recent observations indicate that a
possible value can be $\Omega=1$ \bib{boomerang} and that likely
the correct value of $\Omega_{m}$ is about $1/3$  \bib{supernovae}
and therefore, that about 90\% of this mass is not luminous (see
e.g.\bib{khalil}).

It needs to be stressed that the visual inspection of galaxy's
distribution suggests nothing else that the absence of large
amount of luminous matter in vast regions. Furthermore it is not
very clear whether the voids are physically empty approximately
spherical regions or larger underdense regions with arbitrary
shapes. Many definition of voids  have been proposed, but a
definitive conclusion has not been reached yet \bib{sch}.

Two solutions of the Einstein equations are generally employed to
study the theoretical properties of the cosmological voids. The
Lema\^{i}tre-Tolman (L-T) metric and the Einstein-Straus Swiss
Cheese model. The first is a spherically symmetric metric for
irrotational dust. A complete account of the developments of this
model is discussed in \bib{krasinski}.

The Swiss Cheese model is obtained by cutting out spherical
regions from a Friedmann-Lema\^{i}tre-Robertson-Walker (FLRW)
model, with null pressure, and substituting them with regions with
a spherically symmetric metric such as the Schwarzschild or the
L-T solutions. Appropriate junction conditions have to be imposed
in order to join the solutions of the Einstein equations. A
problem which was solved  by Einstein and Straus \bib{einstraus}
in order to study the effect of the universal expansion on the
planetary orbits (McVittie and J\"arnefeldt studied the same
problem some time before, see references in \bib{einstraus2}).

The L-T and the Swiss Cheese models do not predict the formation
of voids, which must be contained in the initial data
\bib{krasinski}

In a series of papers, Piran and his coworkers attribute the void
formation to the evolution of negative perturbations
\bib{Blumenthal:1991ms} \bib{Dubinski:1992tr}
\bib{Piran:1997fs} \bib{friedpir}. In particular in
\bib{Piran:1997fs} it is shown that these negative fluctuations
behave as if they have a negative effective mass; in
\bib{friedpir} Friedmann and Piran  show that the underdense
regions can result as the combined effect of the gravitational
expansion of negative density perturbations and biasing, because
galaxies are less likely to form in an underdense region.

In this paper, we show that void formation can be the result of
the collapse of positive perturbations. To this purpose, we
consider an Einstein-Straus model with a distribution of spherical
voids of fixed comoving radius $R_{v}$. In the centre of each void
we assume that a black hole, whose mass $M$ compensates the mass
that the void would have if it were completely filled by matter
with the average cosmological mass-energy density $\rho$. In what
follows  we calculate the mass, the Schwarzschild radius and the
densities of these black holes. Due to their cosmological
properties, we shall call them {\sl cosmological black holes}
(CBH).  As they have apparently very low densities, these   CBHs
may have likely formed directly from  the collapse of very large
wavelength perturbations. Therefore voids are the result of the
comoving expansion of the matter surrounding the CBHs.

This scenario explains the existence of a large amount of dark
matter, which is hidden in these black holes without perturbing
the cosmic background radiation. In this simple model the CBHs do
not affect the galaxy motion due to the Birkhoff theorem, they
just take part in the collective universal expansion.

In the last part of the paper we discuss some observational
implications of the proposed scenario, in particular the
gravitational lensing effects induced by these CBHs behave and
their influence on the cosmic background radiation

In the final remarks we discuss the limits of this cosmological
model. For instance we expect to gain some improvement by
weakening the condition of exact sphericity of the CBHs and of the
voids.

\section{ \normalsize Properties and Origin
of the Cosmological Black Holes}\label{esu}

We consider here an Einstein-Straus universe characterised  by the
following properties,

\noindent a) For sake of simplicity we shall limit ourselves to
consider a flat Friedmann-Lema\^{i}tre-Robertson-Walker universe;

\noindent b) in all the voids there is a central spherical black
hole with mass $M$
\begin{equation}\label{massflat}
 M=\frac{4}{3}\pi \Omega_{cbh}\rho_{crit.} R_{v}^{3},
\end{equation}
where the parameter
\begin{equation}\label{omegacbh}
  \Omega_{cbh}=\frac{\rho_{cbh}}{\rho_{crit.}}
\end{equation}
represents the fraction of density due to  all these black holes
to the total density of the universe; where $\rho_{crit.}
=1.88h^{2}\times 10^{-29}\,g\, cm^{-3}$ is the critical density of
the universe;

\noindent and

\noindent c) all the voids are spherical.

We also note that the Schwarzschild radius can be related to the
mass-energy density $c^{2}\rho=\epsilon$, see \bib{chandra2}.

It is known that a  black hole forms when a body with mass $M$
collapses entirely within a sphere of radius

\begin{equation}\label{schw}
R_s=\frac{2GM}{c^2}.
\end{equation}
This statement is equivalent to say that its density satisfies the
relation,
\begin{equation}\label{schwdens1}
R_s(\rho)= \sqrt{\frac{3c^2}{8\pi G\rho}}.
\end{equation}
Conversely,  Eq. (\ref{schwdens1}) relates to any density  a
corresponding Schwarzschild radius. In other words any space-like
sphere of matter with uniform density $\rho$ and radius at least
equal to $R_s(\rho)$ is a black hole.

Whereas astrophysical and primordial black holes form at very high
densities \bib{blantho}, relation (\ref{schwdens1}) implies that
it is possible to consider also the formation of low-density black
holes. The transition of collapsing matter to a black hole at low
densities is described in \bib{smooth2}.

From Eqs. (\ref{massflat}) and (\ref{schw}) we determine the mass
$M$ and the corresponding Schwarzschild radius and consequently
from Eq. (\ref{schwdens1}) the mass-energy density of the central
black hole. In Table 1, we list, in solar units, the results for
three typical diameters of voids.
\begin{table}\label{table}
\begin{tabular}{@{}cccccccc}
   Void diameter& Mass$\times h/\Omega_{cbh}$& $R_{s}\times h/\Omega_{cbh}$ & density \\ \hline
  $30 h^{-1} $ Mpc & $ 3.9\times 10^{15}M_{\odot}$ & $0.37$ kpc&$5.02\times10^{-15}g/cm^{3}$ \\ \hline
  $50h^{-1}$ Mpc& $1.8\times10^{16} M_{\odot}$& $1.7$ kpc& $2.34\times10^{-16}g/cm^{3}$\\ \hline
  $100h^{-1}$  Mpc &$1.4 \times 10^{17} M_{\odot}$ & $13.9$ kpc & $3.66 \times 10^{-18} g/cm^{3}$\\ \hline
\end{tabular}
 \caption{CBHs corresponding to different size voids.}
\end{table}

The low densities given in the Table 1 are those reached by the
collapsing matter when it crossed the Schwarzschild radius.  Loeb
in \bib{loeb}, observes that massive black holes can form from the
collapse of primordial gas clouds after the recombination epoch
($200\leq (1+z) \leq 1400$). We expect that  the process of
formation of a CBH started with large wavelength perturbations at
cosmological densities smaller than $10^{-19}g/cm^{3}$ i.e for
$1+z  \approx 10^{3}$. According to the inflationary scenario, we
only need to suppose that the inflation occurred during a time
long enough to provide such perturbations.

The Oppenheimer-Snyder model \bib{opsny} describes the spherical
symmetric collapse at zero pressure. A growing perturbation, in a
homogeneous and isotropic universe, with initial wavelength
$\lambda_i$ and amplitude $\delta_i <<1$, it first expands and
then collapses according to the equations,
\begin{equation}\label{expans}
\left( \frac{\dot a_{p}}{a_{i}}\right)^{2}=
H^{2}_{i}\left(\Omega_{p}(t_i)\frac{a_{i}}{a_{p}}+1-\Omega_{p}(t_i)\right),
\end{equation}

\begin{equation}\label{omega}
  \Omega_{p}(t_i)=\frac{\rho(t_i)(1+\delta_{i})}{\rho_{c}(t_i)}=\Omega(t_i)(1+\delta_i).
\end{equation}
The solution of Eq. (\ref{expans}) and the relation
\begin{equation}\label{lungdond}
\lambda(t)= \lambda_{i}\frac{a(t)}{a_i}
 \end{equation}
give, in a parametric form, the evolution of $\lambda(t)$,
\begin{equation}\label{a}
\lambda(\theta)=\frac{\lambda_i}{2}
\frac{\Omega_p}{\Omega_p-1}(1-\cos\theta)
\end{equation}
and
\begin{equation}\label{t}
   t(\theta)=
   \frac{1}{2H_i}\frac{\Omega_p}{(\Omega_p-1)^{\frac{3}{2}}}
   (\theta-\sin\theta),
\end{equation}
where $a_i$ is the expansion factor at the beginning of the
perturbation formation, $H_{i}$ is the corresponding Hubble
constant and $\Omega_{p}$ is the ratio of the perturbation density
and the background critical density. It is important to note that,
borrowing a sentence in \bib{einstraus2}, due to spherical
symmetry, it follows that inside the {\sl collapsing matter} and
in a certain neighborhood behaves as if the {\sl collapsing
matter} were embedded in a space ``with no cosmic expansion or
curvature''.

A serious objection to this picture could be that very unlikely
perturbations develop with an exact spherical symmetry. As a
matter of fact, one of the principal processes that can prevent
the collapse of a perturbation into a black hole is the
acquisition of angular momentum through the tidal torques produced
by interaction with other structures. The presence of an angular
momentum produces a centrifugal barrier at typical scales of six
order of magnitude larger than the Schwarzschild radius. But it
has been shown by Loeb \bib{loeb} that, for large perturbations,
the friction between the collapsing matter and the cosmic
background radiation is capable to extract angular momentum and
energy and reduce the centrifugal barrier below the Schwarzschild
radius. We assume that the total mass of the perturbation
\begin{equation}\label{totalmass}
M=\frac{\pi}{6} \Omega_{p}\rho_{i} \lambda_i^{3}
\end{equation}
remains constant during the whole process.

The Schwarzschild radius of the spherical perturbation is equal to

\begin{equation}\label{raggio}
 R_{s}=\frac{H_i^2}{c^2}\left(\frac{\lambda_i}{2}\right)^3\Omega_p.
\end{equation}

We can therefore distinguish two cases. First, the case in which
the relation
\begin{equation}\label{raggiomax}
\frac{2 R_{s}}{\lambda_{i}}\geq1,
 \end{equation}
is satisfied,  the perturbation is in the linear regime  and,
according to the evolution equations in a universe with constant
equation of state, it is frozen when $\lambda_{i}$ is larger than
the Hubble radius \bib{mukhanov}. After crossing the Hubble
horizon, it collapses and becomes a black hole when
\begin{equation}\label{limiti}
\left(\frac{ \lambda_{i}}{2}\right)^{2}\geq
\frac{c^{2}}{H_{i}^{2}\Omega_p}.
\end{equation}
In the second case
\begin{equation}\label{raggiomin}
\frac{2R_{s}}{\lambda_i} <1,
\end{equation}
the perturbation evolves according to Eqs. (\ref{a}) and
(\ref{t}). During the contraction, it becomes unavoidably a black
hole, since the final density is very low and the internal
pressure and temperature can not raise to values large enough to
prevent the collapse, even when the perturbation enters in a
non-linear regime. In addition, any possible centrifugal barrier
is reduced to values smaller than the Schwarzschild radius.

Let us establish a limit for the black hole formation. To this aim
we fix a threshold value $\hat\rho$ of the mass-energy density,
above which one can expect that the equation of state sensibly
changes and then equilibrium conditions can be established, thus
preventing any further collapse, see a similar discussion in
\bib{chandra2}.

To $\hat\rho$ we can associate an final wavelength $\hat\lambda$
defined by the equation
\begin{equation}\label{nocbh1}
M=\frac{\pi}{6}\hat\rho{\hat\lambda}^{3}.
\end{equation}
Comparison with Eq. (\ref{totalmass}) yields
\begin{equation}\label{nocbh2}
 {\hat\lambda}^{3}=\frac{\rho_i\Omega_p}{\hat\rho}\lambda^3_i.
\end{equation}
A perturbation does {\sl not collapse} to a black hole  when
 $\hat\lambda/2>R_{s}$, i.e. if its initial wavelength satisfies
the equation
 \begin{equation}\label{nocbh3}
\left(\frac{ \lambda_i}{2}\right)^{2}< \frac{3c^{2}}{8\pi G}
 \frac{1}{(\rho_i\Omega_p)}
 \left(\frac{\rho_i\Omega_p}{{\hat\rho}}\right)^{\frac{1}{3}}.
\end{equation}

According to the Einstein-Straus model, after the formation of the
black hole, the matter around it expands in a comoving way leading
to the formation of an empty region between it and the rest of the
universe.  As the central black hole cannot be seen, the whole
region appears to an external observer as a void.

In conclusion, among all wavelengths of the cosmological
perturbations spectrum, only those structures which satisfy
(\ref{nocbh3})  appear in the observed universe as {\sl luminous}
matter or {\sl exotic} dark matter. The rest of the matter is
confined in very massive cosmological black holes.

\section{\normalsize Some phenomenological aspects of the CBHs}\label{phenom}

A CBH can be detected through its lensing properties, it must
behave as a Schwarzschild  gravitational lens. Since, according to
our hypothesis, a CBH sits in the centre of a void, the Einstein
angle is \bib{sef}

\begin{equation}\label{einstangle}
 \alpha_{0} =  4.727 \times 10^{-4}\Omega_{CBH}^{1/2}
 \sqrt{ \frac{R_{V}^{3}D_{ds}}{D_{d}D_{s}}}
\end{equation}
where $R_{V}$is the radius of the void, $D_{s}$ is the distance of
the source from the observer, $D_{ds}$ is the distance of the
source from the CBH, and $D_{d}$ is the distance of the CBH from
the observer, all this quantities are expressed in  Mpc. The
characteristic length
\begin{equation}\label{carac}
 \xi_{0}=\alpha_{0}D_{d}.
\end{equation}
For a 50 Mpc void of diameter, with the centre placed at a
distance of 80 Mpc from the Sun and with the source at the
opposite edge of the void, we expect
\[\alpha_{0}\simeq 3.2\times
10^{-3}\Omega_{CBH}^{1/2}
\]
 and
\[
\xi_{0}\simeq 2.6 \times 10^{-1}\Omega_{CBH}^{1/2} Mpc.
\]

This value is the expected extension of an Einstein ring, but to
observe it is required that the source must lie exactly on top of
the resulting degenerate point-like caustic \bib{wamb}; so, in
order to falsify our model, a first point is to analyze the
probability that a galaxy in the background can satisfy this
demand. We expect a general magnification effect in the
neighborhood of the CBH \bib{bontz}. Considering a galaxy with an
effective luminous part long about $6$  kpc , we expect in case of
a perfect alignment a magnification factor of the order of
$1.4\times 10^{2}$. Numerical calculations indicate that the
magnification factor decays to $6.8$ in the range of $6\times
10^{-2}$ Mpc  and reaches the value of $1.37$ at  $0.6$ Mpc .

The high masses involved produce also strong field lensing effects
\bib{ohanian} \bib{bozzaeal} \bib{Vir}, in the range of
distances of about $3R_s$. In the case of a $50$ Mpc  void we
expect, theoretically, the production of an infinite sequence of
{\sl relativistic images}, on both sides of the optical axis, at
scales of the order of $\sim 6$ kpc.

A third observational effect is the influence of a CBH on the
cosmic background radiation. In \bib{zelsa} Zeldovich and Sazhin
point out that generally static structures can raise the
temperature of the cosmic background radiation by an amount
proportional to the Hubble parameter and the gravitational time
delay. By considering the Swiss Cheese model case, they find a
fluctuation of temperature $\delta T/T \sim 10^{-10}$ for a giant
galaxy ($M= 4\times10^{12} M_{\odot}$). As this result is
proportional to the mass, it follows that it corresponds to a
fluctuation of temperature $\delta T/T \sim 10^{-5}$ for a large
CBH with mass ($M\sim 10^{17} M_{\odot})$, which does not
contradict the recent COBE measurements.

Very recent observations of type Ia supernovae showed that the
expansion of the universe is accelerating \bib{supernovae}. This
acceleration  can be justified by admitting that the universe is
dominated by a source of ``dark energy'' with negative pressure.
Constraints on this source require that $\Omega_{dark\
ener.}\simeq 2/3$ and dark and baryonic matter contribute to the
total energy density of the universe with $\Omega_{M}\sim 1/3$
\bib{supernovae2}.

Measurements, performed using  different methods, led to a wide
range of values results of $\Omega_{M}$. According to the
mass-to-light ratio Bahcall and coworkers obtained
$\Omega_{M}(M/L)=0.16 \pm 0.05$
\bib{bahcall1} and recently, after some corrections, $\Omega_{M}(M/L)=0.17 \pm 0.05$
\bib{bahcall2}. It is important to note that according to
\bib{bahcall3} it was assumed that the voids do not contribute
with additional dark matter. On the other side estimates of the
baryon fraction in clusters yield $\Omega_{M}\leq 0.3 \pm 0.05$
and the evolution of cluster abundance gives $\Omega_{M}\simeq
0.25$ (see \bib{bahcall4}).

Finally Turner \bib{turner2001} infers that $\Omega_{M}= 0.33 \pm
0.035$ together with $\Omega_{B}= 0.039 \pm 0.0075$ according to
recent measurements of the physical properties of clusters, CMB
anisotropy and the power spectrum of mass inhomogeneity.

We think that our model could reconcile the discrepancy between
the mass-to-light ratio method and the others which find larger
values of $\Omega_{M}$. As a matter of fact the mass-to-light
ratio cannot trace the presence of the CBHs.

As voids occupy about 50 \% of the volume of the universe, we
claim that the CBHs would simply double the mass observed by the
mass-to-light ratio method, i.e. $\Omega_{CBH}\sim\Omega_{M}(M/L)$
and
\begin{equation}\label{omega1}
  \Omega_{M}=\Omega_{CBH}+\Omega_{M}(M/L)
\end{equation}
This hypothesis can be confirmed by measuring the Einstein angles
produced by the CBHs, according to equation (\ref{einstangle}).

\section{\normalsize Remarks and conclusions}\label{conclusions}

Before concluding this work we remind that the idea that black
holes can be generated by cosmological perturbations has been
already used to predict the existence of  primordial black holes
\bib{blantho}\bib{zelno} \bib{haw71} \bib{carr1}
\bib{carr2}. But as dark matter candidates their role is limited by
very severe constraints \bib{gunn}.

The formation of supermassive black holes from the collapse of a
primordial cloud was proposed by Loeb in
\bib{loeb}. Recent observations provide some evidence of
the existence of supermassive black holes \bib{richstone}. These
objects have been revealed in the centers of the galaxies with
masses up to $10^{9} M_{\odot}$.

In this paper we considered the existence of the CBHs, with larger
masses and residing at the center of the voids, isolated from any
other form of matter. This model explains part of the dark matter
problem and provides new observational predictions, it is possible
to detect the presence of the cosmological black holes by
observing their lensing properties on a background of galaxies.

Moreover we remark that the existence of these black holes is
compatible with the Zeldovich perturbation spectrum and with any
other spectrum. A more precise information on the spectrum can be
obtained by analyzing the void distribution, which reflects the
black hole distribution.

But we think that our model can be improved. First of all it is
based on exact spherical symmetry, which is not a realistic
condition when dealing with the collapse of a perturbation. Second
since the CBHs are generated by perturbations, their mass does not
necessarily correspond to the value rated by (\ref{massflat}),
even if the severe restrictions on the initial amplitudes of the
perturbations, imposed by the COBE observations, allow us to take
the values of mass, energy-mass density, and the Schwarzschild
radius given in Table 1 as good approximations of the real ones.
Third the Einstein-Straus model is not stable with respect to the
variations of mass and to radial perturbations (see
\bib{krasinski} and the references within).

Moreover as stated by the Birkhoff theorem, a spherical CBH, that
compensates a spherical void, does not influence the motions of
the galaxies outside the void. But, according to Peebles in
\bib{peebles}, some peculiarities of the galaxy motion can be
explained by admitting the presence of large amounts of mass in
the voids. This requires that the effective mass of a cosmological
black hole may be larger than the mass given by (\ref{totalmass}).
In this case that the void expansion must not be comoving as in
the Einstein-Straus model (for a review on this problem see
\bib{krasinski}).

The previous considerations and the observation that presently
voids  present an underdense distribution of galaxies (see
\bib{El-Ad:1997af}), suggest  that the CBHs can affect the motion
of the surrounding galaxies. This can be explained only relaxing
the conditions of the Birkhoff theorem and considering the non
uniform distribution of matter on lower scales. This problem will
be analyzed by computer simulations.

Finally we think to develop our model extending the results
obtained from the Einstein-Straus Swiss-Cheese model with the
recent results obtained by Bonnor \bib{bonnor1}\bib{bonnor2},
Senovilla and Vera \bib{senovilla1}, Mars \bib{mars1} \bib{mars2}.

\vskip.5truecm

\noindent{\bf Aknowledgments} I am very grateful to
Prof.~G.~Cosenza, Dr.~G.~Bimonte, Dr.~G.~Esposito and
Dr.~P.~Santorelli for their constant encouragement; I am also
indebted with Prof.~G.~Platania, Dr.~G.~Covone, Dr. C.~Rubano
Dr~P.~Scudellaro for very stimulating discussions and with
Prof.~G.~Longo for important remarks and suggestions that helped
to improve the final version of this work.

This paper is dedicated to the memory of Prof. R.~de Ritis.

\end{document}